\documentclass{iopart}

\usepackage{url}
\usepackage{amsbsy}
\usepackage{graphicx}
\input{epsf} 

\newcommand{\eqref}[1]{{(\ref{#1})}}
\def\bSo{{\hat{\bf S}_1}}
\def\bSt{{\hat{\bf S}_2}}
\def\bL{{\hat{\bf L}_{N}}}
\def\hn{{\hat{\bf n}}}
 
\begin{document}

\title{The Mock LISA Data Challenges: from Challenge 1B to Challenge 3}

\author{The \emph{Mock LISA Data Challenge Task Force}:
Stanislav Babak$^1$,
John G. Baker$^2$,
Matthew J. Benacquista$^3$,
Neil J. Cornish$^4$,
Jeff Crowder$^5$,
Shane L. Larson$^6$,
Eric Plagnol$^7$,
Edward K. Porter$^1$,
Michele Vallisneri$^\mathrm{5,8}$,
Alberto Vecchio$^\mathrm{9}$
and the \emph{Challenge-1B participants}:
Keith Arnaud$^2$,
Leor Barack$^{11}$,
Arkadiusz B{\l}aut$^{12}$,
Curt Cutler$^\mathrm{5,8}$,
Stephen Fairhurst$^{13}$,
Jonathan Gair$^\mathrm{14,1}$,
Xuefei Gong$^{15}$,
Ian Harry$^{13}$,
Deepak Khurana$^{16}$,
Andrzej Kr\'olak$^{17}$,
Ilya Mandel$^\mathrm{8,10}$,
Reinhard Prix$^{18}$,
B. S. Sathyaprakash$^{13}$,
Pavlin Savov$^8$,
Yu Shang$^{15}$,
Miquel Trias$^{19}$,
John Veitch$^9$,
Yan Wang$^{20}$,
Linqing Wen$^{1,8,21}$,
John T. Whelan$^1$}

\address{$^1$ Max-Planck-Institut f\"ur Gravitationsphysik (Albert-Einstein-Institut), Am M\"uhlenberg 1, D-14476 Golm bei Potsdam, Germany}
\address{$^2$ Gravitational Astrophysics Lab., NASA Goddard Space Flight Center, 8800 Greenbelt Rd., Greenbelt, MD 20771, USA}
\address{$^3$ Center for Gravitational Wave Astronomy, Univ.\ of Texas at Brownsville, Brownsville, TX 78520, USA}
\address{$^4$ Dept.\ of Physics, Montana State Univ., Bozeman, MT 59717, USA}
\address{$^5$ Jet Propulsion Laboratory, California Inst.\ of Technology, Pasadena, CA 91109, USA}
\address{$^6$ Dept.\ of Physics, Weber State Univ., 2508 University Circle, Ogden, UT 84408, USA}
\address{$^7$ APC, UMR 7164, Univ.\ Paris 7 Denis Diderot, 10, rue Alice Domon et Leonie Duquet, 75025 Paris Cedex 13, France}
\address{$^8$ Theoretical Astrophysics, California Inst.\ of Technology, Pasadena, CA 91125}
\address{$^9$ School of Physics and Astronomy, Univ.\ of Birmingham, Edgbaston, Birmingham B152TT, UK}
\address{$^{10}$ Dept.\ of Physics and Astronomy, Northwestern Univ., Evanston, IL, USA}
\address{$^{11}$ School of Mathematics, Univ.\ of Southampton, Southampton, SO171BJ, UK}
\address{$^{12}$ Inst.\ of Theoretical Physics, Univ.\ of Wroc{\l}aw, Wroc{\l}aw, Poland}
\address{$^{13}$ School of Physics and Astronomy, Cardiff Univ., 5, The Parade, Cardiff, CF243YB, UK}
\address{$^{14}$ Inst.\ of Astronomy, Univ.\ of Cambridge, Madingley Rd., Cambridge, CB30HA, UK}
\address{$^{15}$ Inst.\ of Appl.\ Math, Acad.\ of Math.\ and System Sci., Chinese Academy of Sciences, 55 Zhongguancun Donglu, Beijing, 100080, China}
\address{$^{16}$ Indian Institute of Technology, Kharagpur, India}
\address{$^{17}$ Inst.\ of Mathematics, Polish Academy of Sciences, Warsaw, Poland}
\address{$^{18}$ Max-Planck-Institut f\"ur Gravitationsphysik (Albert-Einstein-Institut), D-30167 Hannover, Germany}
\address{$^{19}$ Dept. de F\'{\i}sica, Universitat de les Illes Balears, Cra.\ Valldemossa km 7.5, E-07122 Palma de Mallorca, Spain}
\address{$^{20}$ Dept.\ of Astronomy, Nanjing Univ., 22 Hankou Road, Nanjing, 210093, China}
\address{$^{21}$ School of Physics, M013, Univ.\ of Western Australia, 35 Stirling Highway, Crawley, WA 6009, Australia}

\ead{Michele.Vallisneri@jpl.nasa.gov}

\begin{abstract}
The Mock LISA Data Challenges are a programme to demonstrate and encourage the development of LISA data-analysis capabilities, tools and techniques. At the time of this workshop, three rounds of challenges had been completed, and the next was about to start.
In this article we provide a critical analysis of entries to the latest completed round, Challenge 1B. The entries confirm the consolidation of a range of data-analysis techniques for Galactic and massive--black-hole binaries, and they include the first convincing examples of detection and parameter estimation of extreme--mass-ratio inspiral sources.
In this article we also introduce the next round, Challenge 3. Its data sets feature more realistic waveform models (e.g., Galactic binaries may now chirp, and massive--black-hole binaries may precess due to spin interactions), as well as new source classes (bursts from cosmic strings, isotropic stochastic backgrounds) and more complicated nonsymmetric instrument noise.
\end{abstract}

\vspace{-0.5cm}



\section{Introduction}

The Laser Interferometer Space Antenna (LISA), an ESA--NASA mission to survey the gravitational-wave (GW) sky at frequencies between $10^{-5}$ and $10^{-1}$ Hz, will record gravitational radiation from millions of sources, most of them in our Galaxy, but many populating the low-to-high--redshift Universe \cite{lisa}.
Such a variety of signals, overlapping in both the time and frequency domains (to the point of creating \emph{confusion noise} at some frequencies) poses a number of interesting new challenges for GW data analysis, whose solution is essential if we are to draw the greatest possible science payoff from such a bold and innovative observatory.

At the end of 2005, the LISA International Science Team (LIST) initiated a programme of Mock LISA Data Challenges (MLDCs) with the goal of understanding at the conceptual and quantitative level the peculiarities of LISA data analysis, of demonstrating LISA's observational capabilities, and of kickstarting the development of data-analysis algorithms, pipelines, and infrastructural elements.
An MLDC Task Force, chartered by the LIST, periodically issues \emph{challenge} data sets containing GW signals from sources of undisclosed parameters, embedded in synthetic LISA noise; challenge participants have a few months to analyze the data and submit detection candidates, which are then compared with the sources originally injected in the data sets. (\emph{Training} data sets with public source parameters are also provided to help participants tune and troubleshoot their codes.)

Three rounds of MLDCs had been completed at the time of this workshop, each spanning approximately six months. Challenge 1 \cite{mldclisasymp,mldcgwdaw1} was focused on establishing basic techniques to observe GWs from compact Galactic binaries, intrinsically monochromatic, isolated or moderately interfering; as well as from the inspiral phase of bright, isolated, nonspinning massive--black-hole (MBH) binaries. Challenge 2 \cite{mldcgwdaw2,mldcamaldi2} featured three considerably more complex data-analysis problems: a data set containing GW signals from approximately 26 million Galactic binaries (again monochromatic) drawn from a randomized population-synthesis catalog; a data set (the ``whole enchilada'') with a similar Galactic-binary population, plus GW signals from an unknown number (between 4 and 6) of nonspinning-MBH binary and from five extreme--mass-ratio inspirals (EMRI); and five more data sets with single-EMRI signals.

The very steep increase in complexity introduced over a short time-scale with Challenge 2 and the need to consolidate analysis techniques  (especially so for EMRIs) before moving to even more taxing challenges motivated the organization of Challenge 1B, a repeat of Challenge 1 with the addition of single-EMRI data sets. Challenge-1B data sets were distributed in the late summer 2007, with a deadline of December 2007 for entries. Ten collaborations submitted solutions. Highlights from this round include the range of techniques used, the participation of a number of new groups that successfully recovered signals from galactic binaries and MBH binaries, and the first convincing demonstration of EMRI detection \emph{and} parameter estimation. Section \ref{s:challenge-1b} provides a brief summary of the entries; additional details about the work of individual collaborations are given elsewhere in this volume.

As we write (April 2008), Challenge-3 data sets have just been released, with entries due at the beginning of December 2008. Challenge 3 represents a definite step in the direction of more realistic source models (such as chirping Galactic binaries and spinning-MBH binaries) and of new source classes (such as short-lived bursts and stochastic backgrounds). Section \ref{s:challenge-3} describes the Challenge-3 data sets and waveform models in detail.

\section{Report on Challenge 1B}
\label{s:challenge-1b}

Challenge 1B focused on three classes of GW sources, each tackled in a separate subchallenge: monochromatic Galactic binaries, MBH binaries and EMRIs. The Galactic-binary data sets (1B.1.1a--c and 1B.1.2--5) and the MBH-binary data sets (1B.2.1--2) had a duration of approximately one year (31457280 s, sampled at intervals of 15 s), while the EMRI data sets (1B.3.1--5) were twice as long (with the same sampling time). The challenge solutions submitted by the participants were assessed with the simple criteria adopted in previous rounds \cite{mldcgwdaw1, mldcamaldi2}. Detector-response data were generated with the \emph{best-fit} source parameters $\vec{\lambda}_\mathrm{sub}$ submitted by the participants, using the same code previously employed to build the challenge data sets. These data were then compared to the detector response to the \emph{true} waveforms, using as a figure of merit the \emph{recovered} SNR
\begin{equation}
\mathrm{SNR}(\vec{\lambda}_\mathrm{sub}) = \frac{(A_\mathrm{true}|A_\mathrm{sub}) + (E_\mathrm{true}|E_\mathrm{sub})} {\sqrt{(A_\mathrm{sub}|A_\mathrm{sub}) + (E_\mathrm{sub}|E_\mathrm{sub})}}\,,
\label{e:SNR}
\end{equation}
where $A$ and $E$ denote time series for the noise-orthogonal TDI observables $(2X - Y - Z)/3$ and $(Z - Y)/\sqrt{3}$ \cite{sensitivity} and $(\cdot|\cdot)$ denotes the usual signal product weighted by instrument noise.
We also quote the \emph{correlation} $C = \mathrm{SNR} / \mathrm{SNR}_\mathrm{opt}$, where 
$\mathrm{SNR}_\mathrm{opt} = \sqrt{(A_\mathrm{true}|A_\mathrm{true}) + (E_\mathrm{true}|E_\mathrm{true})}$ is the \emph{optimal} SNR. For a perfect detection $C = 1$, but fluctuations $\sim 1 / \mathrm{SNR}_\mathrm{opt}$ are expected because of instrument noise. When we examine parameter errors, these are defined simply as
$\Delta\lambda^i = \lambda^i_\mathrm{sub} - \lambda^i_\mathrm{true}$; in some cases it makes sense to consider the \emph{fractional} parameter errors $\Delta \lambda^i / \lambda^i = (\lambda^i_\mathrm{sub} - \lambda^i_\mathrm{true}) / \lambda^i_\mathrm{true}$.
In the rest of this section we briefly discuss the entries submitted by participants; the technical notes accompanying the entries can be found at \url{www.tapir.caltech.edu/~mldc/results1B/results.html}.

\subsection{Galactic binaries: Challenges 1B.1.X}

Data sets 1B.1.1a--c and 1B.1.2--5 contained GW signals from monochromatic Galactic binaries, in a variety of parameter ranges and source combinations. Seven parameters are required to fully characterize each such source: the amplitude $\mathcal{A}$, the (constant) frequency $f$, the ecliptic latitude and longitude $\beta$ and $\lambda$, the inclination and polarization angles $\iota$ and $\psi$, and the initial phase $\phi_0$. Entries were submitted by five groups: \textbf{GSFC} (scientists at Goddard Space Flight Center); \textbf{IMPAN} (the Institute of Mathematics of the Polish Academy of Science and the Institute of Theoretical Physics at the University of Wroc{\l}aw); \textbf{AEI} (the Albert Einstein Institute in Golm, Germany); \textbf{MCMNJU} (the Institute of Applied Mathematics of the Chinese Academy of Sciences and the Department of Astronomy of Nanjing University); \textbf{UIBBham} (the University of the Balearic Islands and the University of Birmingham). However, all groups except AEI concentrated only on a subset of the challenges.

Participants employed a fair range of techniques, in the same broad class as adopted for similar challenges in the past \cite{mldcgwdaw1,mldcamaldi2}; however, new implementations and different technical solutions are being pursued. \textbf{GSFC} used the X-Ray Spectral Fitting Package (XSPEC) \cite{XSPECwebsite} to fit templates to energy spectra. The package includes a Levenberg--Marquardt optimization algorithm, which was used to obtain an initial guess for the source parameters. A Markov-Chain Monte Carlo (MCMC) routine, also available in XSPEC, was then used to converge to the best-fit source parameters. \textbf{IMPAN} set up a grid-based matched-filtering search with an optimized placement of templates on a hypercubic lattice. The ${\cal F}$-statistic \cite{fstat} was used to reduce the search space from $7$ to $3$ parameters. A similar technique was adopted by \textbf{AEI} \cite{prixwhelan}, in conjunction with a rigid--adiabatic model of detector response. \textbf{MCMNJU} used a genetic algorithm that optimized the ${\cal F}$-statistic; \textbf{UIBBham} implemented a MCMC search described in more detail in \cite{tvv}.
\begin{figure}
\centerline{\includegraphics[width=7.5cm]{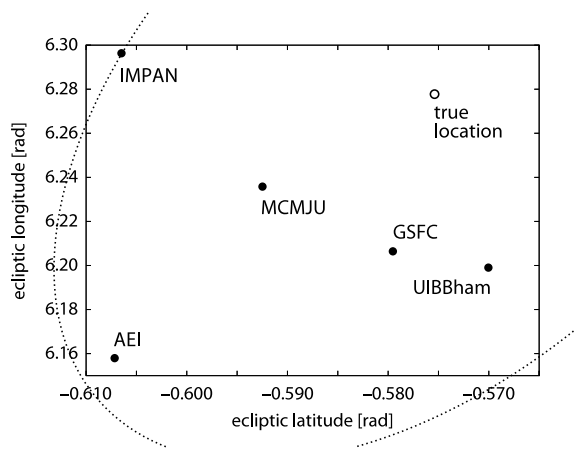}}
\caption{Sky positions reported by participants in Challenge 1B.1.1a, compared to the true location of the source. All the sky positions fall within the Fisher-matrix 1-$\sigma$ contour, shown as a dotted line. (The spread of the reported sky positions is due to both the difference in noise realizations between the Synthetic LISA and LISA Simulator versions of data set 1B.1.1a, and the systematic errors of the searches.)\label{Figure_1b_1_1a_sky_positions}}
\end{figure}

Each of data sets 1B.1.1a--c contained a GW signal from a single monochromatic binary (differing by frequency), with SNR $\approx 13\mbox{--}25$. Table \ref{Table_1b_1_1_correlations} lists the correlations recovered by each collaboration. Some of the entries included close matches for the intrinsic parameters ($f$, $\theta$, and $\phi$), but not for the remaining (extrinsic) parameters: this was due to the (relative) inaccuracy of the LISA response models used by the participants, or by mismatch between their definitions of the extrinsic parameters and the MLDC's. In these cases we recomputed $C$ by maximizing the ${\cal F}$-statistic \cite{fstat} for the intrinsic parameters provided; the resulting $C$s are denoted by asterisks, and show that the intrinsic parameters were indeed recovered well enough to ensure solid detections. 
Table \ref{Table_1b_1_1_parameter_differences} lists parameter errors, and figure \ref{Figure_1b_1_1a_sky_positions} shows where in the sky each collaboration placed the single binary of data set 1B.1.1a, compared to its true position, and to the Fisher-matrix 1-$\sigma$ error contour.
\begin{table}
\caption{Correlations for single--Galactic-binary challenges 1B.1.1a--c. Asterisks denote entries corrected by maximizing the ${\cal F}$-statistic and using the resulting extrinsic parameters; these corrections are not reported where the frequency is well off, and the $\mathcal{F}$-statistic is merely fitting noise.\label{Table_1b_1_1_correlations}}
\begin{indented} \lineup
\item[]\begin{tabular}{llll}
\br
Group & \centre{1}{1B.1.1a}
& \centre{1}{1B.1.1b}
& \centre{1}{1B.1.1c} \\
& \centre{1}{$\mathrm{SNR}_\mathrm{opt}=13.819$}
& \centre{1}{$\mathrm{SNR}_\mathrm{opt}=24.629$}
& \centre{1}{$\mathrm{SNR}_\mathrm{opt}=15.237$} \\
\mr
AEI & $0.108 \rightarrow 0.984^*$ & $0.922 \rightarrow 0.996^*$ & $-0.190 \rightarrow 0.989^*$ \\
GSFC & $0.992$ & $0.807 \rightarrow 0.814^*$ & $-0.138$ \\
IMPAN & $0.988$ & $0.981 \rightarrow 0.997^*$ & \m$0.924 \rightarrow 0.946^*$ \\
MCMNJU & $0.952 \rightarrow 0.996^*$ & $0.906 \rightarrow 0.994^*$ & \m$0.033$ \\
UIBBham & $0.992$ & $0.996$ & \\
\br
\end{tabular}
\end{indented}
\end{table}
\begin{table}
\caption{Parameter errors for Challenges 1B.1.1a--c. All angles are expressed in radians.\label{Table_1b_1_1_parameter_differences}}
\lineup \scriptsize \flushright
\begin{tabular}{llllllll}
\br
Group & \centre{1}{$\Delta \beta$} & \centre{1}{$\Delta \lambda$} & \centre{1}{$\Delta f$ [nHz]} & \centre{1}{$\Delta\psi$} & \centre{1}{$\Delta \iota$} & \centre{1}{$\Delta \varphi$} & \centre{1}{$\Delta \mathcal{A}$ [$10^{-23}$]}\\
\mr
\centre{8}{Challenge 1B.1.1a ($f_\mathrm{true} = 1.060$ mHz)} \\[2pt]
AEI		& $-0.032$	& $-0.120$	& \0\0$-2.43$		& \m0.217		& $-0.454$ 	& \m1.17		& \m1.22		\\
GSFC		& $-0.004$	& $-0.071$	& \0\0$-1.81$	 	& \m0.708	 	& \m0.252 	& \m1.33	 	& \m1.20		\\
IMPAN		& $-0.031$	& \m0.018 	& \0\0\m2.13		& \m0.454 	    & \m0.212	 	& $-1.06$ 	& \m1.25		\\ 
MCMNJU		& $-0.017$	& $-0.042$	& \0\0$-0.53$	& \m0.662		& \m0.426		& $-1.57$		& \m2.34		\\
UIBBham		& \m0.005	& $-0.079$	& \0\0$-1.51$	& \m0.708		& \m0.173		& $-1.32$		& \m0.65		\\
\mr
\centre{8}{Challenge 1B.1.1b ($f_\mathrm{true} = 2.904$ mHz)} \\[2pt]
AEI		& $-0.056$	& $-0.0090$	& \0\m\00.95		& $-1.050$		& \m0.283		& \m1.63		& $-0.066$	\\
GSFC		& $-0.462$		& \m0.0606	& \0$-30.9$		& \m2.560		& \m0.182		& \m0.52		& $-0.024$	\\
IMPAN		& \m0.020 	& \m0.0007	& \0\m\00.85 	& \m0.333 	& \m0.339 	& $-0.60$ 	& \m0.713		\\ 
MCMNJU		& $-0.067$	& $-0.0063$	& \0\m\02.07	& $-0.732$	& $-0.064$	& \m0.84		& $-0.223$	\\
UIBBham		& $-0.044$	& $-0.0082$	& \0\m\01.78	& $-0.636$	& \m0.043	& \m1.13		& $-0.029$	\\
\mr
\centre{8}{Challenge 1B.1.1c ($f_\mathrm{true} = 9.943$ mHz)} \\[2pt]
AEI		& $-0.026$	 & \m0.0053	& \m\0\01.84		& $-0.499$	& $-1.120$		& \m3.02		& \m0.124		\\ 
GSFC		& $-0.452$		& $-1.48$		& \m140		& \m1.820		& $-0.471$	& $-0.66$	& $-0.695$	\\ 
IMPAN		& $-0.016$	& \m0.0248	& \m\0\03.72		& $-1.510$		& $-0.197$	& \m2.68		& \m0.478		\\ 
MCMNJU		& $-0.555$		& $-0.3680$	& \m359		& $-1.590$		& $-0.250$	& $-0.94$	& $-0.532$	\\ 
\br
\end{tabular}
\end{table}

Data set 1B.1.2 contained GW signals from 25 ``verification'' binaries of known (i.e., disclosed) frequency and sky location. Five of them were taken from the list of observed binaries on Gijs Nelemans' wiki \cite{nelemanswiki}, while the remaining 20 were placed randomly in the Galaxy, varying their frequencies over a representative range. Table \ref{Table_1b_1_4_correlations} lists the \emph{global} correlations (computed for the combined signals of all reported and true binaries) recovered by the three groups that participated in this challenge. Just as it happened for Challenges 1B.1.1a--c, problems in assigning extrinsic parameters reduced the correlations. (Since the intrinsic parameters were provided to the participants, we did not perform ${\cal F}$-statistic--based adjustments, which would amount to solving the entire problem. As a result, the low $C$ values of table \ref{Table_1b_1_4_correlations} may be symptomatic only of extrinsic-parameter systematics.)

Data set 1B.1.3 contained GW signals from 20 unknown binaries distributed across the LISA band, well separated in frequencies. Unfortunately, a bug in the random generation of source parameters caused all SNRs to be too small for detection (all were below 1).
Happily, no participating group reported a positive detection, consistent with the correct behavior expected of search algorithms.

\begin{figure}
{\flushright\includegraphics[width=10cm]{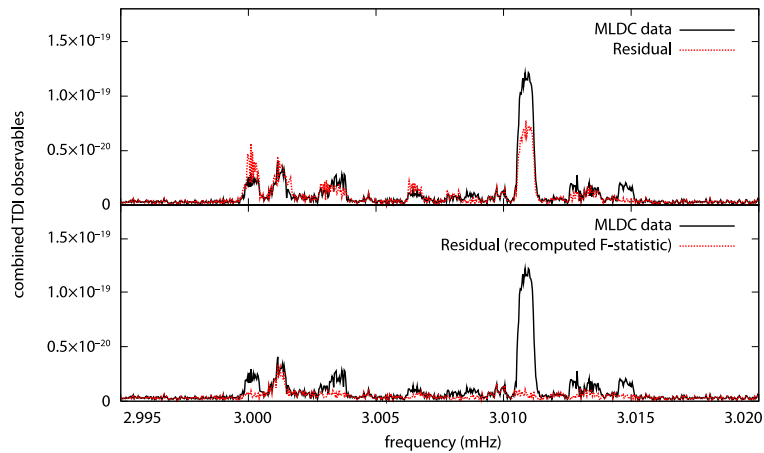}\\}
\caption{Combined $A$ and $E$ spectral amplitude for the GW signals in data set 1B.1.4, before and after subtracting the \textbf{AEI} signal model. In the bottom panel, the extrinsic parameters for all binaries in the model were adjusted by maximizing the ${\cal F}$-statistic.\label{Figure_1b_1_4_AEI}}
\end{figure}

Challenge 1B.1.4 was meant to test search algorithms in the presence of mild source confusion. Fifty-one sources were spread across a band of $15\, \mu{\rm Hz}$ beginning at $3\, {\rm mHz}$, with an average density of 0.108 sources per frequency bin. By contrast, Challenge 1B.1.5 tested algorithms in the presence of a higher level of source confusion, comparable to that expected from our Galaxy. Forty-four sources were spread across a band of $3\, \mu{\rm Hz}$ centered at $3\, {\rm mHz}$, with an average density of 0.465 sources per frequency bin. Table \ref{Table_1b_1_4_correlations} lists the global correlations and the number of sources recovered by the participating groups, as well as the number of \emph{false positives}, defined here as reported sources farther than one frequency bin (1/year) from any true source, or with $\mathcal{F}$-statistic--adjusted correlation less than 0.7 with all true sources within a frequency bin.
The top panel of figure \ref{Figure_1b_1_4_AEI} shows the combined $A$ and $E$ spectral amplitude for the GW signals in data set 1B.1.4, together with the residual after subtracting the signal model submitted by \textbf{AEI}. The bottom panel shows the same subtraction after extrinsic parameters have been recomputed for this entry by maximizing the $\mathcal{F}$-statistic.

Altogether, it must be said that Challenge 2 provided a more forceful demonstration of LISA's science objectives for Galactic binaries \cite{mldcamaldi2}; but this round of challenges was still very useful for new groups to start implementing search methods, and for established groups to continue tuning them. The extrinsic-parameter reporting errors seen here are easy to commit, because these parameters are very sensitive to the modeling of the LISA response, and because their definitions are somewhat conventional, but these errors have little bearing on detection confidence. To avoid such problems in the future, we plan to provide a web tool to check the recovered SNR against the challenge data sets using the fiducial MLDC waveform-generation code.
\begin{table}
\caption{\label{Table_1b_1_4_correlations} Correlations, number of recovered sources, and number of false positives for Challenges 1B.1.2, 1B.1.4, 1B.1.5. Correlations computed after correcting extrinsic parameters using the ${\cal F}$-statistic are denoted with asterisks.}
\begin{indented} \lineup
\item[]\begin{tabular}{lll}
\br
Group & \centre{1}{$C$} & \centre{1}{\# recovered} \\
\mr
\centre{3}{1B.1.2 (${\rm SNR}_{\rm opt}=634.918$, $25$ sources)}  \\[2pt]
AEI		& $-0.822$	& 25		\\
GSFC		& \m0.006	& 25		\\
MCMNJU	& \m0.267	& 25		\\
\mr
\centre{3}{1B.1.4 (${\rm SNR}_{\rm opt}=340.233$, $51$ sources)} \\[2pt]
AEI		& \m$0.774 \rightarrow 	0.966^*$	& 13	 (2 false pos.)	\\
GSFC		& \m$0.003 \rightarrow 	0.282^*$	& \06 (1 false pos.)	\\
\mr
\centre{3}{1B.1.5 (${\rm SNR}_{\rm opt}=273.206$, $44$ sources)} \\[2pt]
AEI		& \m$0.453 \rightarrow 0.929^*$	& \03 \\
\br
\end{tabular}
\end{indented}
\end{table}

\subsection{Massive Black Hole Binary Systems: Challenges 1B.2.X}

Each of data sets 1B.2.1 and 1B.2.2 contained a loud GW signal from a single MBH binary embedded in instrument noise. Gravitational radiation was modeled as the \emph{restricted} waveform for spinless point masses moving on an adabatic sequence of circular orbits, evolving according to 2PN energy-balance equations \cite[sec.\ 4.4]{mldcgwdaw2}. (See also section \ref{sec:ch3mbh} for the new waveform features being introduced for Challenge 3.) Nine parameters are needed to describe each source: the two masses $m_1$ and $m_2$, the time of coalescence $t_c$, the sky-position angles $\beta$ and $\lambda$, the luminosity distance $D_L$, the orbital-inclination and GW-polarization angles $\iota$ and $\psi$, and the initial orbital phase $\varphi_0$. 
The binaries in both data sets had masses drawn from the same ranges ($m_1 = 1\mbox{--}5 \times 10^6\,M_{\odot}$, $m_1/m_2 = 1\mbox{--}4$), but were distinguished by the times of coalescence $t_c = 6\pm 1$ months for Challenge 1B.2.1 and $400 \pm 40$ days (past the end of the data set) for Challenge 1B.2.2; the $\mathrm{SNR}_\mathrm{opt}$ for the sources were chosen to be $\simeq 500$ and $\simeq 80$, respectively.

Two groups submitted entries: \textbf{JPL} (a collaboration between researchers at Caltech and at the Jet Propulsion Laboratory) employed a three-step hierarchical strategy combining a time--frequency track-search analysis, a template-bank matched-filtering search, and a final MCMC stage to evaluate the posterior probability densities of source parameters for data sets 1B.2.1 and 1B.2.2. \textbf{Cardiff} (a collaboration based at that university) used a stochastic--template-bank matched-filtering search \cite{cardiffmbh} to analyze data set 1B.2.1.

Table \ref{mbh} summarizes these entries. While the recovered SNRs are very close to $\mathrm{SNR}_\mathrm{opt}$, which indicates detections of very high confidence, there are large discrepancies in the sky-position angles for Challenge 1B.2.1, where both the \textbf{JPL} and \textbf{Cardiff} searches converged on secondary likelihood maxima, very close in height to the true mode (as shown by the recovered SNR), but quite distant in parameter space.
In fact, the \textbf{JPL} result places the source almost at the antipodal sky position, even if $\mathrm{SNR} \simeq \mathrm{SNR}_\mathrm{opt}$ to better than one part in a thousand.

This is a true global degeneracy, which does not appear in local Fisher-matrix analyses (another example of why mock-data endeavors are useful!), and which may indicate the need, in EM searches of counterparts to LISA binary-MBH detections, to examine unconnected regions of the sky. It may however be premature to make such an inference, since the degeneracy could be broken by spin effects (now included in Challenge-3 waveforms) and higher waveform harmonics. In addition, EM-counterpart searches would require a data-analysis system capable of determining the sky position of MBH binaries a few days in advance of their merger (corresponding to the interval between data dumps from LISA to the ground), whereas data set 1B.2.1 included the inspiral waveform all the way to the approximate merger frequency. Therefore this challenge was \emph{not} aimed directly at establishing the feasibility of sky-position determination for EM-counterpart searches.
In this context, it is however worth pointing out that the entries provided very accurate determinations of the times of coalescence, corresponding to time windows of a few minutes (for Challenge 1B.2.1) and about 45 minutes (for 1B.2.2).

If we can compare the errors of Table 4 with the parameter-determination accuracies predicted in the Fisher-matrix formalism, we see that the \textbf{JPL} estimates for $m_1$ and $m_2$ fall within the 2-$\sigma$ contour for data set 1B.2.1 (notwithstanding the problem with sky position), and near the 1-$\sigma$ contour for 1B.2.2. For data set 1B.2.2, \textbf{JPL}'s $\Delta t_c$, $\Delta D_L$, $\Delta \beta$, and $\Delta \varphi_0$, are also close to 1-$\sigma$, and $\Delta \lambda$ is $\sim 2.2 \sigma$; the errors in $\iota$ and $\psi$ are all tens of $\sigma$s (but the Fisher-matrix formalism is not always reliable for extrinsic parameters). Altogether, we conclude that the \textbf{JPL} search essentially achieves the theoretical limits of parameter extraction for data set 1B.2.2, while it does so for an important subset of parameters for data set 1B.2.1.

\begin{table}
\caption{\label{mbh}Relative/absolute errors for the MBH binaries in Challenges 1B.2.1 and 1B.2.2. All angles are given in radians.
Error estimates have been adjusted to account for two \emph{perfect} symmetries of the waveforms with respect to source parameters:
$\psi\rightarrow\psi+\pi$, and the simultaneous transformation $\psi\rightarrow\psi+\pi/2$, $\varphi_0\rightarrow\varphi_0-\pi/2$.}
\begin{indented}\lineup
\item[]\begin{tabular}{@{}llll}
\br
& \multicolumn{2}{c}{1B.2.1 (${\rm SNR}_{\rm opt}=531.84$)} & \multicolumn{1}{c}{1B.2.2 (${\rm SNR}_{\rm opt}=80.67$)} \\
Group          & \multicolumn{1}{c}{JPL} & \multicolumn{1}{c}{Cardiff} & \multicolumn{1}{c}{JPL} \\
\mr
${\rm SNR}$ 	& \m531.57	& \m511.78 & \0\0\m79.86	\\
$\Delta m_{1}/m_{1}$  & \0\0\m$5.991 \times10^{-3}$   & \0\0\m0.108    & \0\0\0\m0.122\\
$\Delta m_{2}/ m_{2}$ & \0\0$-5.252 \times10^{-3}$    & \0\0$-0.111$    & \0\0\0$-0.134$\\
$\Delta t_c$ [s]     & \m206.1  & $-541.8$ & $-2688.2$ \\
$\Delta D_L/ D_L$     & \0\0$-0.139$                    & \0\0$-1.438$         & \0\0\0\m$4.781\times10^{-3}$\\
$\Delta \beta$        & \0\0\m2.429                   & \0\0\m1.374         & \0\0\0\m$5.862\times10^{-3}$\\
$\Delta \lambda$      & \0\0\m3.133                   & \0\0\m0.548          & \0\0\0$-1.461\times10^{-2}$\\
$\Delta \iota$        & \0\0\m0.713                   & \0\0\m0.678         & \0\0\0$-6.955\times10^{-2}$\\
$\Delta \psi$         & \0\0$-0.564$                     & \0\0\m1.448 & \0\0\0$-4.878\times10^{-2}$\\
$\Delta \varphi_0$    & \0\0$-2.846$                    & \0\0$-2.389$        & \0\0\0\m$1.293 \times 10^{-2}$ \\
\br
\end{tabular}
\end{indented}
\end{table}

\subsection{Extreme Mass Ratio Inspirals: Challenges 1B.3.X}

Although Challenge 2 saw a few successful detections of EMRI signals \cite{mldcamaldi2}, data sets 1B.3.1--5 represented the first real testbed for the search algorithms developed for this critical source class. Each data set contained a GW signal from a single EMRI embedded in instrument noise, with $\mathrm{SNR}_\mathrm{opt}$ between $\simeq$ 55 and $\simeq$ 135, and source parameters chosen randomly as described in \cite{mldcgwdaw2}. Fourteen parameters are needed to describe each EMRI source \cite{mldcgwdaw2}: 
the ecliptic latitude and longitude $\beta$ and $\lambda$ and the luminosity distance $D_L$; the central-BH and compact-object masses $M$ and $\mu$;  the magnitude $a$ and orientation angles $\theta_K$, $\phi_K$ of the central-BH spin; the initial radial orbital frequency $\nu_0$ and eccentricity $e_0$; and three angles $\tilde{\gamma}_0$, $\alpha_0$ and $\lambda$ describing the initial orientation of the orbit.

Entries were received from three groups: \textbf{BBGP} (a collaboration of scientists at the AEI, Cambridge, and the University of Southampton); \textbf{EtfAG} (AEI, Northwestern, and Cambridge); and \textbf{MT} (Montana State University).
\textbf{EtfAG} employed a time--frequency technique \cite{gmw}, whereas \textbf{BBGP} \cite{bbgpemri} and \textbf{MT} \cite{cornishemri} developed coherent approaches based on Monte Carlo techniques, differing in their implementation.
The entries were assessed as discussed at the beginning of this section, although the \textbf{EtfAG} time--frequency analysis cannot determine the extrinsic parameters, so their recovered SNR and correlation could not be computed. Tables \ref{EMRI2} and \ref{EMRI1} summarize all results.

Both the time--frequency and coherent approaches succeeded in detecting these relatively strong EMRI signals and in constraining their parameters (with especially remarkable accuracy for data set 1B.3.1), although not all groups analyzed all data sets, and the performance of the same search pipeline varied across them. Challenge participants report that in some cases this was due to a lack of time for extended computations before the challenge deadline, so some parameter set were submitted as ``best fits'' although they were clearly understood to be secondary likelihood maxima. Thus, this early development work indicates that the main challenge for (isolated) EMRI analyses is the very complex structure of the likelihood surface in source-parameter space, which features a number of secondary maxima of similar height, even more so than for MBH binaries.

We caution the reader that it would be inappropriate at this time to draw general conclusions about the relative merits of search methods and about the expected science payoff of LISA EMRI astronomy: it is not known how these techniques scale as the SNR decreases and in situations where EMRI signals overlap with each other and are affected by Galactic confusion noise. The first two complications will be addressed in Challenge 3.
\begin{table}
\caption{\label{EMRI2} Overlaps and recovered SNRs for TDI observables $A$, $E$ and combined recovered SNR for data sets 1B.3.1--5.}
\begin{indented}\lineup
\item[]\begin{tabular}{@{}llllll}
\br
Group & \centre{1}{$C_A$} & \centre{1}{$\mathrm{SNR}_A$} &  \centre{1}{$C_E$} & \centre{1}{$\mathrm{SNR}_E$} & \centre{1}{total SNR} \\
\mr
	 \centre{6}{1B.3.1 ($\mathrm{SNR}_\mathrm{opt} = 123.7$)}	\\[2pt]
BBGP & \m0.57 & \m51.0 & \m0.58 & \m51.6 & \m\072.5\\
MT  & \m0.998 & \m86.1 & \m0.997 & \m88.3 & \m123.4\\
\mr
	 \centre{6}{1B.3.2 ($\mathrm{SNR}_\mathrm{opt} = 133.5$)}	\\[2pt]
BBGP & \m0.07 & \m\06.6 & \m0.18 & \m18.2 & \m\017.6\\
BBGP$^\mathrm{a}$ & \m0.39 & \m37.6 & \m0.41 & \m39.8 & \m\054.7\\
MT  & \m0.54 & \m49.5 & \m0.54 & \m50.8 & \m\070.9 \\
\mr
	 \centre{6}{1B.3.3 ($\mathrm{SNR}_\mathrm{opt} = 81.0$)}	\\[2pt]
BBGP & $-0.06$ & $-4.2$ & $-0.0003$ & $-0.05$ & \0$-3.0$ \\
BBGP$^\mathrm{a,c}$ & $-0.2$ & $-11.5$ & $-0.32$ & $-19.0$ & \0$-21.5$\\
MT  & \m0.38 &  \m22.0 & \m0.35 & \m20.9 & \m\030.4 \\
\mr
	 \centre{6}{1B.3.4 ($\mathrm{SNR}_\mathrm{opt} = 104.5$)}	\\[2pt]
BBGP$^\mathrm{c}$ & \m0.0007 & \m\02.1 & $-0.0002$ & $-0.8$ & \m\02.1\\
BBGP$^\mathrm{b}$ & \m0.16 & \m13.9 & \m0.04 & \m6.7 & \m\014.6 \\ 
\mr
	 \centre{6}{1B.3.5 ($\mathrm{SNR}_\mathrm{opt} = 57.6$)}	\\[2pt]
BBGP & \m0.09 & \m\03.4 & \m0.1 & \m4.2 & \m\05.3\\
\br
\end{tabular}
\item[]$^\mathrm{a}$ $C$ and SNR after correcting the sign of $\beta$, lost on input to the MLDC webform.
\item[]$^\mathrm{b}$ $C$ and SNR after correcting phases at $t=0$, to account for a \textbf{BBGP} bug.
\item[]$^\mathrm{c}$ The \textbf{BBGP} SNRs can be negative because \textbf{BBGP} maximized likelihood analytically over amplitude, which makes SNR sign-insensitive (a minus sign corresponds to a change of $\pi$ in the phase of the dominant harmonic). This degeneracy is broken when all the harmonics are found correctly.
\end{indented}
\end{table}
\begin{table}
\caption{Errors for a subset of EMRI parameters in Challenges 1B.3.1--5. ``$\Delta x/[x]$'' denote fractional errors relative to the physical or prior range of the parameter.
Note that due to the narrow prior range the relative error could be limited to $\sim 10\%$,
therefore we give the number which would show by how much we have decreased the prior range. Large errors correspond to the secondary maxima in the likelihood, the only true 
parameters (global maximum) were found by \textbf{MT} in Challenge 1.3.1.
\label{EMRI1}}
\scriptsize\lineup
\begin{tabular}{l@{\hspace{3pt}}l@{\hspace{3pt}}l@{\hspace{3pt}}l@{\hspace{3pt}}l@{\hspace{3pt}}l@{\hspace{3pt}}l@{\hspace{3pt}}l@{\hspace{3pt}}l@{\hspace{3pt}}l@{\hspace{3pt}}l}
\br
Group &
\centre{1}{$\frac{\Delta\beta}{[\beta]}$} & 
\centre{1}{$\frac{\Delta\lambda}{[\lambda]}$} &
\centre{1}{$\frac{\Delta\theta_K}{[\theta_K]}$} &
\centre{1}{$\frac{\Delta\phi_K}{[\phi_K]}$} &
\centre{1}{$\frac{\Delta a}{[a]}$} &
\centre{1}{$\frac{\Delta\mu}{[\mu]}$} & 
\centre{1}{$\frac{\Delta M}{[M]}$} &
\centre{1}{$\frac{\Delta \nu_0}{\nu_0}$} &
\centre{1}{$\frac{\Delta e_0}{0.15}$} & 
\centre{1}{$\frac{\Delta\lambda_{SL}}{[\lambda_{SL}]}$} \\
\mr
\centre{11}{Challenge 1B.3.1}	\\[2pt]
BBGP   & $-0.03$   &   $-0.0059$   &   $-0.14$   &   \m0.053   &   \m0.31   &  $-0.20$   &   $-0.84$   &   \m0.026    &   \m0.37     &   $-0.022$   \\
EtfAG  & \m0.019   &   $-0.0045$   &   \m0.56   &   \m0.33   &   \m0.16   &   $-0.11$   &   $-0.27$   &   $-9.3 \times 10^{-5}$    &   \m0.17     &   \m0.078    \\
MT     &  \m0.0058   &   \m0.0027   &   \m$4.4 \times 10^{-4}$   &   \m0.0051   &   $-0.0022$   &   \m0.0065   &   \m0.014   &   \m$3.2\times 10^{-6}$      &   $-0.0085$    &   $-0.0020$   \\
\mr
\centre{11}{Challenge 1B.3.2}	\\[2pt]
BBGP   &  $-0.16$   &   $-0.43$   &   \m0.46   &   $-0.33$   &   $-0.0088$   &   $-0.0040$   &   \m0.016   &   \m$1.4 \times 10^{-4}$     &   $-0.010$    &   $-0.0013$   \\
EtfAG  &  $-0.014$   &   \m0.0042   &   \m0.97   &   $-0.36$   &   \m0.0043   &   $-0.046$   &   $-0.069$   &   $-6.5\times 10^{-5}$     &   \m0.041     &   \m0.0041  \\
MT     & \m0.0040   &   $-0.0086$   &   \m0.79   &   \m0.41   &   \m0.093   &   $-0.064$   &   \m0.35   &   $-0.035$    &   \m0.068    &   \m0.092    \\
\mr
\centre{11}{Challenge 1B.3.3}	\\[2pt]
BBGP   &  \m0.091   &   \m0.50   &   $-0.23$   &   \m0.045   &   $-0.32$   &   $-0.49$   &   $-0.029$   &   \m$6.1 \times 10^{-4}$      &   \m0.019     &   \m0.054   \\
EtfAG  &  $-0.01$   &   $-0.004$   &   \m0.49   &   $-0.34$   &   \m0.0073   &   $-0.059$   &   $-0.061$   &   $-7.8\times 10^{-5}$    &   \m0.038      &   \m0.0061  \\
MT    &  \m0.045   &   $-0.019$   &   $-0.1$   &   \m0.077   &   $-0.066$   &   \m0.13   &   \m0.59   &   \m$3.6 \times 10^{-4}$   &   $-0.33$     &   \m0.010  \\
\mr
\centre{11}{Challenge 1B.3.4}	\\[2pt]
BBGP   & $-0.57$   &   $-0.37$   &   \m0.37   &   $-0.31$   &   $-0.025$   &   \m0.020   &   $-0.88$   &   \m0.066     &   \m0.065     &   $-0.16$   \\
EtfAG  & $-0.56$   &   \m0.49   &   \m0.56   &   $-0.34$   &   \m0.059   &   \m0.12   &   \m0.04   &   \m$2.8 \times 10^{-4}$    &   $-0.039$    &   \m0.0040   \\
\mr
\centre{11}{Challenge 1B.3.5}	\\[2pt]
BBGP   & $-0.48$   &   $-0.14$   &   $-0.35$   &   \m0.1   &   $-0.094$   &   $-0.094$   &   \m0.55   &   $-0.0021$    &   $-0.017$      &   $-0.060$  \\
EtfAG  & $-0.58$   &   \m0.46   &   \m0.27   &   $-0.084$   &   \m0.20   &   $-0.7$   &   \m0.83   &   $-0.066$     &   \m0.066     &   \m0.27  \\
\br
\end{tabular}
\end{table}

\section{Synopsis of Challenge 3}
\label{s:challenge-3}

The third round of the MLDCs consists of five challenges (3.1--3.5). Data sets 3.1--3 consist of approximately two years of data ($2^{22}$ samples at a cadence of $15$ s) for time-delay interferometry (TDI) observables $X$, $Y$, and $Z$. These data sets are released both as time series of equivalent strain generated by the LISA Simulator \cite{lisasimulator} and as time series of fractional frequency fluctuations generated by Synthetic LISA \cite{synthlisa}; see \cite[p.\ S556]{mldcgwdaw2} for the conversion between the two. Indeed (with a few exceptions, described below, for 3.4 and 3.5), the Challenge-3 data sets are built using the ``pseudo-LISA'' model of Challenges 1 and 2: the orbits of the LISA spacecraft are $e^2$-accurate Keplerian ellipses with conventional orientations and time offsets; \emph{modified} TDI (a.k.a.\ TDI 1.5) expressions are used for the observables; and Gaussian, stationary instrument noise is included from six proof masses and six optical benches with known noise levels that are identical across each set of six.\footnote{The six proof-mass noises are uncorrelated and white in acceleration, with one-sided power spectral density (PSD)
\begin{displaymath}
S_\mathrm{acc}^{1/2}(f) = 3 \times 10^{-15} [1 + (10^{-4}\,{\rm Hz}/f)^2]^{1/2} \, \mathrm{m}\, \mathrm{s}^{-2}\, \mathrm{Hz}^{-1/2};
\end{displaymath}
the six optical-path noises are uncorrelated and white in phase with PSD
\begin{displaymath}
S_\mathrm{opt}^{1/2}(f) = 20 \times 10^{-12} \, \mathrm{m}\, \mathrm{Hz}^{-1/2};
\end{displaymath}
the conversion to Synthetic LISA's dimensionless fractional frequency fluctuations is described on \cite[p.\ 6]{synthlisa}; the values actually used in the MLDCs are
\begin{displaymath}
S_\mathrm{acc}(f) = 2.5 \times 10^{-48} (f/\mathrm{Hz})^{-2} [1 + (10^{-4}\,{\rm Hz}/f)^2] \, \mathrm{Hz}^{-1};
\end{displaymath}
\begin{displaymath}
S_\mathrm{opt}(f) = 1.8 \times 10^{-37} (f/\mathrm{Hz})^{2} \, \mathrm{Hz}^{-1}.
\end{displaymath}
} See \cite{mldcgwdaw2} for details.
\begin{itemize}
\item \textit{Data set 3.1} contains a Galactic GW foreground from $\sim$ 60 million compact binary systems.
This data set is a direct descendant of Challenge 2.1, but it improves on the realism of the latter by including both detached and interacting binaries with intrinsic frequency drifts (either positive or negative). Section \ref{sec:ch3galaxy} gives details about the binary waveform models, about their implementation in the LISAtools suite \cite{lisatools}, and about the generation of the Galactic population. 
\item \textit{Data set 3.2} contains GW signals from 4--6 binaries of spinning MBHs, on top of a confusion Galactic-binary background. This data set improves on the realism of Challenges 1.2.1--2 and 2.2 by modeling the orbital precession (and ensuing GW modulations) due to spin--orbit and spin--spin interactions. Section \ref{sec:ch3mbh} gives details about the MBH-binary waveforms.

Because this challenge focuses on the effects of spins rather than on the joint search for MBH signals and for the brightest Galactic binaries, the background is already \emph{partially subtracted}---it is generated from the population of detached binaries used for Challenge 3.1, withholding all signals with SNR $> 5$.
\item \textit{Data set 3.3} contains five GW signals from EMRIs. As in Challenges 1.3.1--5, EMRI waveforms are modeled with Barack and Cutler's ``analytic kludge'' waveforms \cite{barackcutler}; this challenge introduces the complication of detecting five such signals with lower SNRs, and \emph{in the same data set}. By contrast, Galactic confusion is not included. See section \ref{sec:ch3emri} for details.
\end{itemize}
Challenges 3.4 and 3.5 address the detection of two GW sources that are \emph{new} to the MLDCs, and that have (respectively) bursty and stochastic characters: thus, these searches require an accurate characterization of instrument noise, which in reality will not be available \emph{a priori}, but will be obtained from the LISA measurements themselves.
To model this problem, in data sets 3.4 and 3.5 the levels of the six + six secondary noises have been independently randomized by $\pm 20 \%$; the noises are however still uncorrelated. In addition, these data sets contain time series for all twelve ``raw'' LISA phase measurements $y_{ijk}$ and $z_{ijk}$ \cite{synthlisa}, so that contestants may now build additional TDI observables to help characterize instrument noise. The phase measurements \emph{do} include laser phase noise, because otherwise they would convey extra information unavailable from the real LISA; but laser noise is reduced in level to $\sim$ ten times the secondary noise at 1 mHz, so that it can be canceled relatively easily with TDI 1.5 implemented with moderate timing precision.
To wit:
\begin{itemize}
\item \textit{Data set 3.4} consists of $2^{21}$ samples at a cadence of 1 s ($\sim$ 24 days), and it contains GW burst signals from cosmic string cusps, occurring as a Poissonian random process throughout the data set, with a mean of five events. Details about the waveforms are given in section \ref{sec:ch3string}. The data set is provided only as fractional frequency fluctuations generated by Synthetic LISA.
\item \textit{Data set 3.5} consists of $2^{20}$ samples at a cadence of 2 s (again $\sim$ 24 days), and it contains a stochastic GW background, which is isotropic, unpolarized, Gaussian and stationary; its spectrum grows at low frequencies as $1/f^3$, and its magnitude is set to a few times the secondary noise over a broad range of frequencies. Details about the synthesis of the background and the simpler model of the LISA orbits used for this challenge are given in section \ref{sec:ch3background}. The data set is provided as fractional frequency fluctuations generated by Synthetic LISA and by the new simulator LISACode \cite{lisacode}, recently integrated into the LISAtools suite \cite{lisatools}; thus, cross checks are possible between the two simulators.
\end{itemize}
LISACode \cite{lisacode} was developed at APC-Paris with the purpose of accurately mapping the impact of the different LISA subsystems on its science observations, and of bridging the gap between the basic principles of the LISA measurement and a future, more sophisticated end-to-end simulator. Thus, LISACode includes realistic representations of most of the ingredients that will influence LISA's sensitivity (such as orbits, instrument noise, ultra-stable--oscillator time stamps, phasemeter response functions), internal waveform generators for several kinds of sources (monochromatic and chirping binaries, stochastic backgrounds, etc.), as well as the the possibility to build various TDI combinations.
Many user-defined parameters make it possible to study the impact of different LISA configurations on its sensitivity. LISACode's conventions follow closely those of the MLDCs and of Synthetic LISA.

All the Challenge-3 data sets can be downloaded at \url{astrogravs.nasa.gov/docs/mldc/round3/datasets.html}, encoded in lisaXML \cite{mldclisasymp}, an XML-based format that can be displayed directly in modern web browsers, and handled easily in C/C++, Python, and MATLAB with the LISAtools I/O libraries \cite{lisatools}. Each data set is released in the blind challenge version and in a training version that includes the source parameters used to generate it. Additional training data sets can be generated easily with the LISAtools suite.\footnote{After installing LISAtools following the instructions at \url{code.google.com/p/lisatools/wiki/Install}, generating a training set is as simple as running (say, for Challenge 3.1) 
\begin{displaymath}
\mbox{\texttt{MLDCpipelines2/bin/challenge3.py -T -R 3.1}}
\end{displaymath}}

The remainder of this section describes the GW signal models adopted for each data set. 
See \cite{mldcgwdaw2} for the conversion of the GW polarizations in source frame (given here) to the LISA frame. Table \ref{tab:parameters} is a glossary of source parameters with their symbols and lisaXML descriptors, while table \ref{table:MLDC3} is a summary of the GW content of each data set along with the ranges used to choose source parameters randomly.
\begin{table}
\caption{Source parameters in Challenge 3.\label{tab:parameters}}
\small
\begin{tabular}{llll}
\br
{Parameter} &
{Symbol} &
{Standard parameter name} &
{Standard unit} \\
& & (lisaXML descriptor) & (lisaXML descr.) \\
\mr
\multicolumn{4}{c}{\textit{Common parameters}} \\
Ecliptic latitude   & $\beta$   & \texttt{EclipticLatitude}  & \texttt{Radian} \\
Ecliptic longitude  & $\lambda$ & \texttt{EclipticLongitude} & \texttt{Radian} \\
Polarization angle  & $\psi$    & \texttt{Polarization}      & \texttt{Radian} \\
Inclination         & $\iota$   & \texttt{Inclination}       & \texttt{Radian} \\
Luminosity distance$^\mathrm{a}$ & $D_L$       & \texttt{Distance}          & \texttt{Parsec} \\
\mr
\multicolumn{4}{c}{\textit{Galactic binaries}} \\
Amplitude$^\mathrm{b}$ & $\mathcal{A}$ & \texttt{Amplitude}    & \texttt{1} (GW strain) \\
Frequency           & $f$           & \texttt{Frequency}    & \texttt{Hertz} \\
Frequency derivative  & $\dot{f}$           & \texttt{FrequencyDerivative}    & \texttt{Hertz/Second} \\
Initial GW phase    & $\phi_0$      & \texttt{InitialPhase} & \texttt{Radian} \\
\mr
\multicolumn{4}{c}{\textit{Spinning massive black-hole binaries}} \\
Masses of component MBHs & $m_1$, $m_2$ & \texttt{Mass1}, \texttt{Mass2} & 	\texttt{SolarMass}\\
Magnitude of spins $S_1$, $S_2$ & $a_1$, $a_2$ & \texttt{Spin1}, \texttt{Spin2} & \texttt{MassSquared} \\
Initial orientation of spin $S_1$ & $\theta_{S1},\phi_{S1}$ & \texttt{PolarAngleOfSpin1} &  \texttt{Radian}\\
                                  &                         & \texttt{AzimuthalAngleOfSpin1} &	
\texttt{Radian}\\
Initial orientation of spin $S_2$ & $\theta_{S2},\phi_{S2}$ & \ldots\textit{likewise} &  \\
Time to coalescence & $T_c$ & \texttt{CoalescenceTime}	 &	\texttt{Second}\\
Phase at coalescence & $\Phi_c$ & \texttt{PhaseAtCoalescence}	 &	\texttt{Radian}\\
Initial orientation & $\theta_L$, $\phi_L$ & \texttt{InitialPolarAngleL}	 &	\texttt{Radian} \\
\multicolumn{1}{r}{of orbital momentum} &  & \texttt{InitialAzimuthalAngleL}	 &	\texttt{Radian}\\
\mr
\multicolumn{4}{c}{\textit{EMRIs: see table 5 of \cite{mldcgwdaw2}}} \\
\mr
\multicolumn{4}{c}{\textit{Cosmic string cusp bursts}} \\
Amplitude$^\mathrm{b}$ (Fourier) & $\mathcal{A}$ & \texttt{Amplitude}    & \verb|Hertz^(1/3)| \\
Central time of arrival & $t_C$ & \texttt{CentralTime}    & \texttt{Second} \\
Maximum frequency$^\mathrm{c}$ & $f_\mathrm{max}$ & \texttt{MaximumFrequency}    & \texttt{Hertz} \\
\mr
\multicolumn{4}{c}{\textit{Isotropic stochastic background}} \\
PSD$^\mathrm{b,d}$ at 1 Hz  & $S_h$ & \texttt{PowerSpectralDensity} & \verb|(f/Hz)^-3/Hz| \\
\br
\end{tabular} \\
$^\mathrm{a}$~We do not deal explicitly with the redshifting of sources at cosmological distances, so $D_L$ is a \emph{luminosity} distance, and all masses and frequencies are measured at the Solar-system barycenter and red/blue-shifted by factors $(1+z)^{\pm 1}$ with respect to those measured locally near the sources. \\
$^\mathrm{b}$~Replaces $D_L$ for Galactic binaries, cosmic-string--cusp bursts, and stochastic-background pseudosources. \\
$^\mathrm{c}$~Effectively replaces $\iota$ for cosmic-string--cusp bursts. \\
$^\mathrm{d}$~Note also that $S_h = S_h^\mathrm{tot}/384$; $\psi$ is set to 0, and $\iota$ not used.
\end{table}
\begin{table}
\caption{Summary of data-set content and source-parameter selection in Challenge 3.
Parameters are sampled randomly from uniform distributions across the ranges given below, and all angular parameters (including spin and orbital--angular-momentum directions for MBH binaries) are drawn randomly from uniform distributions over the entire appropriate ranges.
Source distances are set from individual-source SNRs, which are drawn randomly from the ranges specified below (in Challenge 3, ``SNR'' refers to the multiple--TDI-observable SNR approximated as $\sqrt{2} \times \mathrm{max} \{\textrm{SNR}_X,\textrm{SNR}_Y,\textrm{SNR}_Z\}$).
The MBH time of coalescence $t_c$ and the cosmic-string--cusp burst central time $t_C$ are given relative to the beginning of the relevant data sets. \label{table:MLDC3}}
\small
\lineup
\begin{tabular}{l@{\hspace{6pt}}l@{\hspace{6pt}}l}
\br
Data set & Sources & Parameters \\
\mr
\textbf{3.1}
& \textit{Galactic-binary background} & randomized population (see section \ref{sec:ch3galaxy}) \\
& & $\sim 34 \times 10^6$ interacting, $\sim 26 \times 10^6$ detached \\[3pt]
& plus 20 \textit{verification binaries} & known parameters (see section \ref{sec:ch3galaxy}) \\
\mr
\textbf{3.2}
& 4--6 \textit{MBH binaries} & for each: $m_1 = 1\mbox{--}5 \times 10^6\,M_\odot$, $m_1/m_2 = 1\mbox{--}4$, \\
& & $a_1/m_1 = 0\mbox{--}1, a_2/m_2 = 0\mbox{--}1$ \\[3pt]
& \multicolumn{1}{r}{\ldots including} & $\textsc{mbh}_1$: $t_c = \090 \pm \030$ days, $\textrm{SNR} \sim 2000$ \\
& & $\textsc{mbh}_2$: $t_c = 765 \pm \015$ days, $\textrm{SNR} \sim 20$ \\
& \multicolumn{1}{r}{\ldots and 2--4 chosen from} & $\textsc{mbh}_3$: $t_c = 450 \pm 270$ days, $\textrm{SNR} \sim 1000$ \\
& & $\textsc{mbh}_4$: $t_c = 450 \pm 270$ days, $\textrm{SNR} \sim 200$ \\
& & $\textsc{mbh}_5$: $t_c = 540 \pm \045$ days, $\textrm{SNR} \sim 100$\\
& & $\textsc{mbh}_6$: $t_c = 825 \pm \015$ days, $\textrm{SNR} \sim 10$ \\[3pt]
& plus \textit{Galactic confusion} & randomized population with approx.\ $\textrm{SNR} < 5$ \\
& & $\sim 26 \times 10^6$ binaries; no verification \\
\mr
\textbf{3.3} & 5 \textit{EMRIs} & for each: $\mu = 9.5\mbox{--}10.5 \, M_\odot$, $S = 0.5\mbox{--}0.7 \, M^2$, \\
&                                             & time at plunge $= 2^{21}\mbox{--}2^{22} \times 15$ s, \\
&                                             & ecc.\ at plunge $= 0.15\mbox{--}0.25$, SNR $= 10\mbox{--}50$ \\[3pt]
&\multicolumn{1}{r}{\ldots including}         & $\textsc{emri}_1$: $M = 0.95\mbox{--}1.05 \times 10^7 M_\odot$ \\
&& $\textsc{emri}_2$ and $\textsc{emri}_3$: $M = 4.75\mbox{--}5.25 \times 10^6 M_\odot$ \\
&& $\textsc{emri}_4$ and $\textsc{emri}_5$: $M = 0.95\mbox{--}1.05 \times 10^6 M_\odot$ \\
\mr
\textbf{3.4} & \textit{n Cosmic-string--cusp bursts} & (with $n$ Poisson-distributed with mean 5) \\
&                                             & $f_\mathrm{max} = 10^{-3\mbox{--}1} \, \mathrm{Hz}$, $t_C = 0\mbox{--}2^{21}$ s, $\textrm{SNR} = 10\mbox{--}100$ \\
&                                             & all instrument noise levels randomized $\pm 20\%$ \\
\mr
\textbf{3.5} & \textit{Isotropic stochastic background} & $2 \times 192$ incoherent $h_+$ and $h_\times$ sources over sky \\
&                                             & $S^\mathrm{tot}_h = 0.7\mbox{--}1.3 \times 10^{-47} (f/\mathrm{Hz})^{-3} \, \mathrm{Hz}^{-1}$ \\
&                                             & all instrument noise levels randomized $\pm 20\%$ \\
\br
\end{tabular}
\end{table}

\subsection{Chirping Galactic binaries}
\label{sec:ch3galaxy}

Data set 3.1 contains GWs from a population of $\sim 26 \times 10^6$ detached and $\sim 34 \times 10^6$ interacting Galactic binaries. Each binary is modelled as a system of two point masses $m_1$ and $m_2$ in circular orbit with linearly increasing or decreasing frequency (depending on whether gravitational radiation or equilibrium mass transfer is dominant). The polarization amplitudes at the Solar-system barycenter, expressed in the source frame, are given by
\begin{eqnarray}
h^S_+(t) & = & \mathcal{A} \left(1 + \cos^2{\iota}\right) \cos[2\pi (f t + \dot{f} t^2 / 2) + \phi_0], \\
h^S_\times(t) & = & -2 \mathcal{A} (\cos{\iota}) \sin[2\pi (f t + \dot{f} t^2 / 2) + \phi_0], \nonumber
\end{eqnarray}
where the amplitude is derived from the physical parameters of the source as $\mathcal{A} = (2 \mu / D_L) (\pi M f)^{2/3}$, with $M = m_1 + m_2$ the total mass, $\mu = m_1 m_2 / M $ the reduced mass, and $D_L$ the distance; $\dot{f}$ is the (constant) frequency derivative, and $\phi_0$ is the phase at $t = 0$.

Since it would be unfeasible to process millions of barycentric binary waveforms individually through the LISA simulators to compute the TDI-observable time series, we adopt a fast frequency-domain method \cite{Cornish:2007if} that rewrites the LISA phase measurements as the fast--slow decomposition
\begin{equation}
y_{ij}(t) = C(t) \cos(2 \pi f_0 t) + S(t) \sin(2 \pi f_0 t) ;
\end{equation}
the functions $C(t)$ and $S(t)$ describe slowly varying effects such as the rotation of the LISA arms, the Doppler shift induced by orbital motion, and the intrinsic frequency evolution of the source. These ``slow'' terms can be sampled very sparsely and Fourier-transformed numerically, while the ``fast'' sine and cosine terms can be Fourier-transformed analytically. The results are then convolved to produce the LISA phase measurements, and these are assembled into the desired TDI variables. This algorithm is three to four orders of magnitude faster than the time-domain LISA simulators, although it effectively approximates LISA as a rigidly rotating triangle with equal and constant armlengths. See \cite{Cornish:2007if} for full details, and directory \url{MLDCwaveforms/Galaxy3} in LISAtools for the source code.

\begin{figure}
\centerline{\includegraphics[width=10cm]{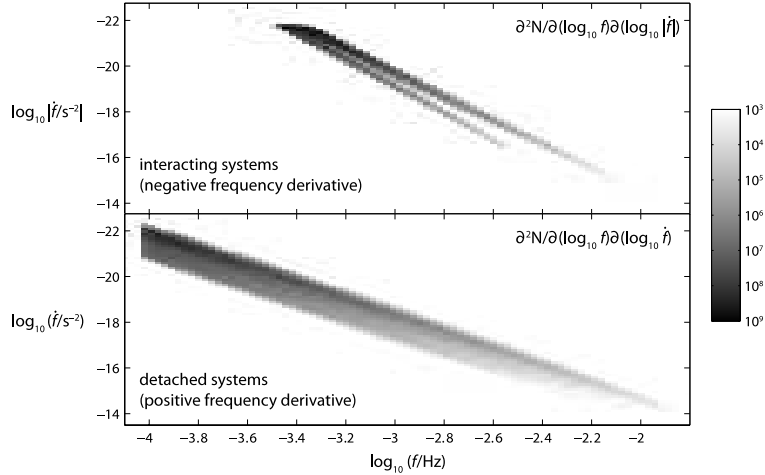}}
\caption{Histogram of the density of Galactic binaries in the Nelemans catalogs, binned by $\log_{10} f$ and $\log_{10} |\dot{f}|$.\label{fig:binaries}}
\end{figure}

The starting point for each realization of data set 3.1 are two large catalogs provided by Gijs Nelemans (files \url{MLDCwaveforms/Galaxy3/Data/AMCVn_GWR_MLDC.dat} and \url{dwd_GWR_MLDC.dat} in the LISAtools installation), which contain the parameters of $26.1 \times 10^6$ detached and $34.2 \times 10^6$ interacting systems produced by the population synthesis codes described in \cite{Nelemans:2001hp, Nelemans:2003ha}. Figure \ref{fig:binaries} shows the distribution of the binaries in the catalogs over $f$ and $\dot{f}$. Recent work by Roelofs, Nelemans and Groot \cite{Roelofs:2007rn} suggests that the model in \cite{Nelemans:2003ha} overpredicts the number of (AM CVn) interacting systems by a factor of 5--10, but we did not implement this correction for Challenge 3. 

The parameters of each binary in the catalogs are modified by randomly tweaking $f$ by $\pm 1\%$, $A$ by $\pm 10\%$, $\beta$ and $\lambda$ by $\pm 0.5\, {}^\circ$, and by randomly assigning $\psi$, $\iota$, and $\phi_0$ ($\dot{f}$ is computed from the catalog's binary-period derivative and from the tweaked $f$). These random perturbations are large enough to render the original population files useless as answer keys, but small enough to preserve the overall parameter distributions. Binaries with approximate single-Michelson SNR $> 10$ are regarded as ``bright'' and listed in a separate table in the challenge keys. Data set 3.1 includes also 20 verification binaries of known parameters (specified in LISAtools file \url{MLDCwaveforms/Galaxy3/Data/Verification.dat} as rows of $f$, $\dot{f}$, $\beta$, $\lambda$, $A$).

\subsection{Spinning MBH binaries}
\label{sec:ch3mbh}

The spinning-MBH--binary GW signals of data set 3.2 are modeled as restricted waveforms (no higher harmonics) from 2PN circular adiabatic inspirals, with uncoupled orbital frequency evolution and spin and orbital precession. Both the orbital phase and frequency are computed as explicit functions of time, corresponding to \emph{T3} waveforms in the classification of \cite{DIS}:
\begin{eqnarray} \fl
M\omega = \frac1{8}\tau^{-3/8} \biggl[1 &+& \left(\frac{743}{2688} + \frac{11}{32}\eta\right)\tau^{-1/4} - 
            \frac3{10}\left(\pi - \frac{\beta}{4}\right)\tau^{-3/8}  \nonumber \\ 
            &+& \left.
            \left(\frac{1855099}{14450688} + 
           \frac{56975}{258048}\eta + \frac{371}{2048}\eta^2 - \frac{3}{64}\sigma\right)\tau^{-1/2}\right],
\end{eqnarray}
where $M = m_1 + m_2$ is the total mass, $\eta = m_1 m_2/M^2 \equiv \mu/M$ is the symmetric mass ratio, and 
\begin{eqnarray}
\tau = \frac{\eta}{5M}(T_c - t),
\end{eqnarray}
\begin{eqnarray}
\beta &=& \frac1{12}\sum_{i=1,2} \left[ 
\chi_i \left(\bL\cdot{\hat{\bf S}_i}\right)\left(113\frac{m_i^2}{M^2} + 75\eta\right)
\right], \\
\sigma &=& -\frac{1}{48}\eta\chi_1\chi_2\left[ 247(\bSo\cdot\bSt) - 721(\bL\cdot\bSo)(\bL\cdot\bSt)
\right].
\end{eqnarray}
Here $\bL$, ${\bf \hat{S}}_1$ and ${\bf \hat{S}}_2$ 
are the unit vectors along the leading-order angular orbital momentum and the MBH spins. The \emph{intrinsic} orbital phase is 
\begin{eqnarray} \fl
\Phi_{orb} = \Phi_C - \frac{\tau^{5/8}}{\eta}\biggl[ 1 &+& 
  \left(  \frac{3715}{8064} + \frac{55}{96}\eta \right)\tau^{-1/4}
  - \frac{3}{16}(4\pi - \beta) \nonumber \\
  & + & \left. \left( \frac{9275495}{14450688}+ \frac{284875}{258048}\eta+ \frac{1855}{2048}\eta^2 - \frac{15}{64}\sigma  \right) \tau^{-1/2}
\right]; \label{OrbPhN}
\end{eqnarray}
however, because the spin--orbit coupling causes the orbital angular momentum to precess around the total angular momentum, 
the phase that enters the gravitational waveforms contains an additional correction \cite{ACST}:
\begin{equation}
\dot{\Phi} = \omega + \frac{(\bL\cdot\hn)[\bL\times \hn]\cdot\dot{\bf \hat{L}}_N}{1-(\bL\cdot\hn)^2}
\equiv \omega + \delta\dot{\Phi},\label{totPhi}
\end{equation}
where $\hn$ is direction to the source.
The constant of integration in this equation can be redefined so that $\delta\Phi = 0$ at $t=0$.  The equations of precession for $\bL$, ${\bf \hat{S}}_1$ and ${\bf \hat{S}}_2$ are given by (2.9)--(2.11) in 
 \cite{LangHughes}. In the source frame the gravitational polarizations are then given (with respect to a time-varying polarization basis) by
\begin{eqnarray}
h_{+} &=& -\frac{2\mu}{D}(1 + \cos^2 i)(M\omega)^{2/3}\cos{2\Phi}, \nonumber \\
h_{\times} &=& \frac{4\mu}{D}\cos i \,(M\omega)^{2/3}\sin{2\Phi},
\label{eq:polmbh}
\end{eqnarray}
where $\cos{i} = (\bL \cdot \hn)$.
The polarizations $h^S_+$ and $h^S_\times$ in a fixed polarization basis are obtained by way of a rotation by the instantaneous polarization angle
\begin{equation}
\tan{\psi} = \frac{\sin{\beta}\cos{(\lambda -\phi_L)}\sin{\theta_L} - \cos{\theta_L}\cos{\beta}}
{\cos{\beta}\sin{(\lambda - \phi_L)}},
\end{equation}
(where $\theta_L$ and $\phi_L$ define the direction of ${\bf L}_N$) yielding
\begin{eqnarray}
h_{+}^S &=& -h_{+}\cos{2\psi} - h_{\times}\sin{2\psi},\\
h_{\times}^S &=& h_{+}\sin{2\psi} - h_{\times}\cos{2\psi}.
\end{eqnarray}
The end of the inspiral is handled with the exponential taper used also for the MBH-binary waveforms of Challenge 2 \cite{mldcgwdaw2}.
See directory \url{MLDCwaveforms/FastBBH} in LISAtools for the source code for these waveforms.

Data set 3.2 includes also a Galactic confusion background generated from the same \emph{detached-binary} population as used in Challenge 3.1 (interacting systems have typically very small chirp masses and are not expected to make a significant contribution), but withholding all binaries with individual $A + E$ SNR $> 5$, relative to instrument noise plus an estimate of confusion noise, which was derived using a BIC criterion for the resolvability of individual Galactic binaries \cite{Cornish:2007if}:
\begin{equation} \fl
S_{X,\mathrm{gal}} = 16 \, x^2 \sin^2 x \, \mathrm{Hz}^{-1} \times \left\{ \begin{array}{l@{}l@{\;}l@{\;}l}
10^{-44.62} &(f/\mathrm{Hz})^{-2.3}  & \mathrm{for} \, f \in [10^{-4}  ,10^{-3}  ] & \mathrm{Hz}, \\ 
10^{-50.92} &(f/\mathrm{Hz})^{-4.4}  & \mathrm{for} \, f \in [10^{-3}  ,10^{-2.7}] & \mathrm{Hz}, \\
10^{-62.8}  &(f/\mathrm{Hz})^{-8.8}  & \mathrm{for} \, f \in [10^{-2.7},10^{-2.4}] & \mathrm{Hz}, \\
10^{-89.68} &(f/\mathrm{Hz})^{-20.0} & \mathrm{for} \, f \in [10^{-2.4},10^{-2.0}] & \mathrm{Hz}
\end{array}\right.
\label{eq:confusionback}
\end{equation}
(fractional frequency fluctuations, with $x = 2 \pi f L$, $L \simeq 16.6782$ s). The resulting confusion background is consistent with \eqref{eq:confusionback}, which is also used in Challenge 3 (on top of instrument noise) to define the SNRs of GW signals from MBH binaries, EMRIs (Challenge 3.3) and cosmic-string cusps (Challenge 3.4).

\subsection{EMRIs}
\label{sec:ch3emri}

The EMRI waveforms of data set 3.3 are the Barack--Cutler \cite{barackcutler} ``analytic kludges'' used for Challenge 1.3.1--5 and described in \cite[sec.\ 4.5]{mldcgwdaw2}, with the single change in that the number of eccentric-orbit harmonics included in the waveform does not evolve with eccentricity, but is fixed at five (lisaXML parameter \texttt{FixHarmonics}; a value of zero will reproduce the old behavior). See directory \url{MLDCwaveforms/EMRI} in LISAtools for the source code.

\subsection{Cosmic string cusps}
\label{sec:ch3string}

Data set 3.4 contains a number of bursts from cosmic strings, the first of two new GW sources introduced with Challenge 3. Cosmic strings are linear topological defects that may be formed in early Universe at the phase transitions predicted in many elementary-particle and superstring models. Cosmic-string oscillations emit gravitational radiation, with a substantial part of the emission from \emph{cusps}, which can achieve very large Lorentz boosts \cite{cusp1}.
In the limit where the tip of a cusp is moving directly
toward the observer, the observed metric perturbation is a linearly polarized GW with \cite{cusp2}
\begin{equation} \fl
h(t) = A \vert t - t_C \vert^{1/3} \times (\mathrm{incomplete} \; \Gamma \; \mathrm{function} \; \mathrm{envelope}) , \quad
A \sim \frac{G \mu L^{2/3}}{D_L}; \label{cusp}
\end{equation}
%
here $t_C$ is the burst's central time of arrival, 
$G$ is Newton's constant, $\mu$ is the string's mass per unit length, $D_L$
is the luminosity distance to the source, and $L$
is the size of the feature that produces the cusp (e.g., the length of
a cosmic string loop). If the observer's line of sight does not coincide
with the cusp's direction of motion, the waveform becomes a much more
complicated mixture of polarizations~\cite{cusp3}. If the viewing angle $\alpha$ departs
only slightly from zero (which is our assumption), the waveform remains dominantly linearly
polarized, and the sharp spike in \eqref{cusp} is rounded
off, introducing an exponential suppression of Fourier-domain power for frequencies above $f_{\rm max} = 2 / (\alpha^3 L)$.

Following the model used by the LIGO Science Collaboration, we define our cusp waveforms
in the Fourier domain according to
\begin{equation}
|h_+(f)| = {\cal A} f^{-4/3} \left(1 + (f_{\rm low}/f)^2\right)^{-4}, \quad h_\times = 0,
\end{equation}
with $\exp(1 - f/f_\mathrm{max})$ suppression above $f_\mathrm{max}$. The amplitude ${\cal A}$ has dimensions ${\rm Hz}^{1/3}$; $f_{\rm low}$ sets the low-frequency cutoff of what is effectively a fourth-order Butterworth filter, which prevents dynamic-range issues
with the inverse Fourier transforms (for Challenge 3 we set $f_{\rm low} = 1 \times 10^{-5}$ Hz).
The phase of the waveform is set to $\exp \rmi(\pi - 2 \pi f t_C)$ before inverse-Fourier transforming to the time domain. See directory \url{MLDCwaveforms/CosmicStringCusp} in LISAtools for the source code.

\subsection{Stochastic background}
\label{sec:ch3background}

Data set 3.5 contains the second GW source new to Challenge 3: an isotropic, unpolarized, Gaussian and stationary stochastic background. Allen and Romano \cite{stochastic} define a stochastic background as the ``gravitational radiation produced by an extremely large number of weak, independent, and unresolved gravity-wave sources, [...] stochastic in the sense that it can be characterized only statistically.'' Such backgrounds are usually characterized by the dimensionless quantity
\begin{equation}
\Omega_\mathrm{gw}(f) = \frac{1}{\rho_\mathrm{crit}} \frac{\rmd \rho_\mathrm{gw}}{\rmd \log f},
\end{equation}
with $\rho_\mathrm{gw}$ the energy density in GWs, and $\rho_\mathrm{crit} = 3 c^2 H_0^2 / (8 \pi G)$ the closure energy density of the Universe, and they are idealized as the collective, incoherent radiation of uncorrelated infinitesimal emitters distributed across the sky. If the background is isotropic, unpolarized, Gaussian and stationary, the Fourier amplitude $\tilde{h}_A(f,\hat{\Omega})$ of each emitter (with $A$ indexing the  $+$ and $\times$ polarizations, and $\hat{\Omega}$ the direction on the two-sphere) is completely characterized by the power-spectral-density relation \cite{stochastic}
\begin{equation} \fl
\big\langle \tilde{h}^*_A(f,\hat{\Omega}) \tilde{h}_{A'}(f',\hat{\Omega}') \big\rangle =
\frac{3 H_0^2}{32 \pi^3}
|f|^{-3} \Omega_\mathrm{gw}(|f|)
\times \delta_{AA'} \delta(f - f') \delta^2(\hat{\Omega},\hat{\Omega}').
\label{eq:pseudosourcepsd}
\end{equation}
In Challenge 3, we assume a constant $\Omega_\mathrm{gw}(f)$, as appropriate for the primordial background predicted in simple cosmological scenarios. We implement the uncorrelated emitters as a collection of 192 pseudosources distributed at \emph{HEALPix} pixel centers across the sky. HEALPix (the Hierarchical Equal Area isoLatitude Pixelization of spherical surfaces \cite{healpix}) is often used to represent cosmic microwave background data sets; 192 pixels correspond to a twice-refined HEALPix grid with $N_\mathrm{side} = 2^2$.

Each pseudosource consists of uncorrelated pseudorandom processes for $h_+$ and $h_\times$, generated as white noise in the time domain, and filtered to achieve the $f^{-3}$ spectrum of \eqref{eq:pseudosourcepsd}, using the the recursive $1/f^2$ filtering algorithm proposed by Plaszczynski \cite{filtering}, extended to spectral slope $-3$. The algorithm employs a chain of $1/f^2$ infinite--impulse-response filters to reshape the white noise spectrum between minimum and maximum frequencies $f_\mathrm{low}$ and $f_\mathrm{knee}$, set to $10^{-5}$ and $10^{-2}$ Hz in this Challenge (see the source file \url{MLDCwaveforms/Stochastic.py} in LISAtools for the Synthetic LISA implementation).

The \emph{one-sided} PSD of each single-polarization random process (which represents the finite area of a pixel in the sky) is then given by $S_h(f)/2 = 3 H_0^2/(32 \pi^3) f^{-3} \Omega_\mathrm{gw} \times (4 \pi / 192)$.
In data set 3.5, we define $S^\mathrm{tot}_h = (192 \times 2) S_h$ and we set it so that, in the TDI observables, the GW background is a few times stronger than LISA's secondary instrument noise. Namely, 
\begin{equation}
S^\mathrm{tot}_h(f) = 0.7\mbox{--}1.3 \times 10^{-47} (f/\mathrm{Hz})^{-3} \, \mathrm{Hz}^{-1}
\end{equation}
(taking $H_0 = 70 \, \mathrm{km} / \mathrm{s} / \mathrm{Mpc}$, this corresponds to $\Omega_\mathrm{gw}=8.95\times 10^{-12}\mbox{--}1.66\times 10^{-11}$).

One of the more promising approaches to detect GW backgrounds with LISA relies on estimating instrument noise levels by way of symmetrized TDI observables that are insensitive to GWs at low frequencies in the LISA band \cite{zetapaper,hoganbender,rrv}. For realistic LISA orbits, however, the low-frequency behavior of such observables becomes more complicated than discussed in the literature. To simplify the initial investigation of the background-detection problem in data set 3.5, we have therefore approximated 
LISA as a rigidly rotating triangle with equal and constant armlengths (i.e., Synthetic LISA's \url{CircularRotating}).

\section{Conclusion}

Since their inception, the Mock LISA Data Challenges have received remarkable support from the GW community, and have set the stage for many practical demonstrations of the feasibility of LISA's exciting science with present-day data-analysis techniques. Future challenges will feature ever more realistic models of waveforms and instrument noise, and they will endeavour to scope out all important aspects of the LISA science objectives.
In addition, the software tools developed for the MLDCs \cite{lisatools} can be used to generate data sets for many other data-analysis experiments outside the main challenges; and the MLDC \emph{standard model} of LISA's observations (including the MLDC ``pseudo-LISA'' and GW models) is proving extremely valuable to the current analytical investigations of the LISA science performance that are being run by the LIST.

To obtain more information and to participate in the MLDCs, see the official MLDC website (\url{astrogravs.nasa.gov/docs/mldc}) and the Task Force wiki (\url{www.tapir.caltech.edu/listwg1b}).

\ack

SB and EP were supported by the Deutsches Zentrum f\"ur Luft- und Raumfahrt; LW's work by the Alexander von Humboldt Foundation's Sofja Kovalevskaja Programme funded by the German Federal Ministry of Education and Research. 
JG thanks the Royal Society for support and the Albert Einstein
Institute for hospitality and support while part of this work was
being completed. 
SF was supported by the Royal Society, BSS and IH by the UK Science and Technology Facilities Council. 
MV was supported by the LISA Mission Science Office and by JPL's HRDF.
CC's, JC's and MV's work was carried out at the Jet Propulsion Laboratory, California Institute of Technology, under contract with the National Aeronautics and Space Administration. 
MT acknowledges support from the Spanish Ministerio de Eduaci\'on y
Ciencia Research projects FPA-2007-60220, CSD207-00042 and the Govern
de les Illes Balears, Conselleria d'Economia, Hisenda i Innovaci\'o. 
IM was partially supported by NASA ATP Grant NNX07AH22G to Northwestern University. 


\section*{References}

\begin{thebibliography}{99}
%
\bibitem{lisa} Bender P, Danzmann P and the LISA Study Team 1998 ``Laser Interferometer Space Antenna for the Detection of Gravitational Waves, Pre-Phase A Report'' \textbf{MPQ 233} (Garching: Max-Planck-Instit\"ut f\"ur Quantenoptik) 
%
\bibitem{mldclisasymp} Arnaud K A et al. (the MLDC Task Force) 2006 \textit{Laser Interferometer Space Antenna: 6th International LISA Symp. (Greenbelt, MD, 19--23 Jun 2006)} ed Merkowitz S M and Livas J C (Melville, NY: AIP) p 619; \textit{ibid} p 625 (long version with lisaXML description: \textit{Preprint} gr-qc/0609106)
%
\bibitem{mldcgwdaw1} Arnaud K A et al. (the MLDC Task Force and Challenge 1 participants) 2007 \textit{Class. Quant. Grav.} \textbf{24} S529
%
\bibitem{mldcgwdaw2} Arnaud K A et al. (the MLDC Task Force) 2007 \textit{Class. Quant. Grav.} \textbf{24} S551
%
\bibitem{mldcamaldi2} Babak S et al. (the MLDC Task Force and Challenge 2 Participants) 2008 \textit{Class. Quant. Grav.} \textbf{25} 114037
%
\bibitem{sensitivity} Vallisneri M, Crowder J and Tinto M 2008 \textit{Class. Quant. Grav.} \textbf{25} 065005
%
\bibitem{XSPECwebsite} Xspec, an X-ray Spectral Fitting Package, \url{heasarc.gsfc.nasa.gov/docs/xanadu/xspec}
%
\bibitem{fstat} Jaranowski P, Krolak A and Schutz B F 1998 \textit{Phys. Rev. D} \textbf{58} 063001
%
\bibitem{prixwhelan} Whelan J T, Prix R and Khurana D 2008 ``Mock LISA Data Challenge 1B: improved search for galactic white dwarf binaries using an F-statistic template bank'' in this volume (\textit{Preprint} arXiv:0805.1972)
%
\bibitem{tvv} Trias M, Vecchio M and Veitch J 2008
``Markov Chain Monte Carlo searches for Galactic binaries in MLDC-1B data sets'' in this volume (\textit{Preprint} arXiv:0804.4029)
%
\bibitem{nelemanswiki} Gijs Nelemans' LISA wiki, \url{www.astro.kun.nl/~nelemans/dokuwiki}
%
\bibitem{cardiffmbh} Harry I W, Fairhurst S and Sathyaprakash B S 2008 ``A hierarchical search for gravitational waves from supermassive black hole binary mergers'' in this volume (\textit{Preprint} arXiv:0804.3274)
%
\bibitem{gmw} Gair J R, Mandel I and Wen L
``Improved time-frequency analysis of extreme-mass-ratio inspiral signals in mock LISA data'' in this volume (\textit{Preprint} arXiv:0804.1084)
%
\bibitem{bbgpemri} 
Gair J R, Babak S, Porter E K and Barack L 2008 ``A Constrained Metropolis-Hastings Search for EMRIs in the Mock LISA Data Challenge 1B'' in this volume (\textit{Preprint} arXiv:0804.3322)
%
\bibitem{cornishemri} Cornish N J 2008 ``Detection Strategies for Extreme Mass Ratio Inspirals'' in this volume (\textit{Preprint} arXiv:0804.3323)
%
\bibitem{lisatools} LISAtools website and SVN repository, \url{lisatools.googlecode.com}
%
\bibitem{barackcutler} Barack L and Cutler C 2004 \textit{Phys. Rev. D} \textbf{69} 082005
%
\bibitem{lisasimulator} Cornish N J and Rubbo L J 2003 \emph{Phys. Rev. D} \textbf{67} 022001; erratum-ibid. 029905; LISA Simulator website, \url{www.physics.montana.edu/lisa}; now included in \cite{lisatools}
%
\bibitem{synthlisa} Vallisneri M 2005 \emph{Phys. Rev. D} \textbf{71}; Synthetic LISA website, \url{www.vallis.org/syntheticlisa}; now included in \cite{lisatools}
%
\bibitem{lisacode} Petiteau A, Auger G, Halloin H, Jeannin O, Plagnol E, Pireaux S, Regimbeau T and Vinet J--Y 2008 \emph{Phys. Rev. D} \textbf{77} 023002; LISACode website, \url{www.apc.univ-paris7.fr/LISA-France/analyse.phtml}; now included in \cite{lisatools}
%
\bibitem{Cornish:2007if} Cornish N J and Littenberg T B 2007 \textit{Phys. Rev. D} \textbf{76} 083006
%
\bibitem{Nelemans:2001hp} Nelemans G, Yungelson L R and Portegies Zwart S F 2001, \textit{Astron. Astrophys.} \textbf{375} 890
%
\bibitem{Nelemans:2003ha} Nelemans G, Yungelson L R and Portegies Zwart S F 2004, \emph{Mon. Not. Roy. Astron. Soc.} {\bf 349} 181
%
\bibitem{Roelofs:2007rn} Roelofs G H A, Nelemans G and Groot P J 2007
\emph{Mon. Not. Roy. Astron. Soc.} {\bf 382} 685
%
\bibitem{DIS} Damour T, Iyer B and Sathyaprakash B S 2001 \textit{Phys. Rev. D} {\bf 63} 044023
%
\bibitem{ACST} Apostolatos T A, Cutler C, Sussman G J and Thorne K S 1994 \textit{Phys. Rev. D} \textbf{49} 6274
%
\bibitem{LangHughes} Lang R N and Hughes S A 2006 \textit{Phys. Rev. D} {\bf 74} 122001
%
%
\bibitem{cusp1} Damour T and Vilenkin A 2000 \textit{Phys. Rev. Lett.} {\bf 85} 3761
%
\bibitem{cusp2} Siemens X, Creighton J, Maor I, Ray Majumder S,
Cannon K and Read J 2006 \textit{Phys. Rev. D} {\bf 73} 105001
%
\bibitem{cusp3} Siemens X and Olum K D 2003 \textit{Phys. Rev. D} {\bf 68} 085017
%
\bibitem{stochastic} Allen B and Romano J D 1999 \textit{Phys. Rev. D} {\bf 59} 102001
%
\bibitem{healpix} G\'orski K M et al. 2005 \textit{Astrophys. J.} {\bf 622} 759; HEALPix website, \url{healpix.jpl.nasa.gov}
%
\bibitem{filtering} Plaszczynski S 2007 \textit{Fluctuation Noise Lett.} {\bf 7} R1
%
\bibitem{zetapaper} Tinto M, Armstrong J W and Estabrook F B 2000 \textit{Phys. Rev. D} \textbf{63} 021101
%
\bibitem{hoganbender} Hogan, C J and Bender P L 2001 \textit{Phys. Rev. D} \textbf{64} 062002 
%
\bibitem{rrv} Robinson E L, Romano J D and Vecchio A 2008 ``Search for a stochastic gravitational-wave signal in the second round of the Mock LISA Data Challenges'' in this volume (\textit{Preprint} arXiv:0804.4144)
%
\end{thebibliography}
\end{document}